\documentclass
[12pt]{article}
\usepackage{amssymb}
\usepackage{epsfig}

\catcode`\@=11
\def\numberbysection{\@addtoreset{equation}{section}
        \def\theequation{\thesection.\arabic{equation}}}

\def\beq{\begin{equation}}
\def\eeq{\end{equation}}
\numberbysection
\begin{document}
\begin{titlepage}
\begin{center}
\hfill  \\
\vskip 1.in {\Large \bf On the area of the sphere in a Snyder geometry} \vskip 0.5in P. Valtancoli
\\[.2in]
{\em Dipartimento di Fisica, Polo Scientifico Universit\'a di Firenze \\
and INFN, Sezione di Firenze (Italy)\\
Via G. Sansone 1, 50019 Sesto Fiorentino, Italy}
\end{center}
\vskip .5in
\begin{abstract}
We compute the area of a generic $d$-sphere in a Snyder geometry.
\end{abstract}
\medskip
\end{titlepage}
\pagenumbering{arabic}
\section{Introduction}

The quantization of gravity is expected to involve the discretization of some geometric quantities. From heuristic arguments ( based on Ehrenfest principle) Bekenstein \cite{1}-\cite{2} proposed that the area of the event horizon must have a discrete spectrum:

\beq A_n \ = \ 4 \pi r^2 \ \sim \ l^2_p \ n \ \ \ \ \ \ \ n = 1,2,... \label{11} \eeq

where $l_p$ is the Planck length.

Among the possible candidates for quantum gravity we can mention the theories with minimal length. These are the subject of several studies, and among these Snyder algebra stands out. This necessarily leads to a non-trivial deformation of quantum mechanics. Evidence of this is the fact that the symmetry group of the path integral ( the canonical transformations ) is modified by the presence of the minimal length \cite{3}.

In this article we want to highlight another equally important aspect of the Snyder space, which connects to the Bekenstein conjecture of the horizon area of the black hole.
In \cite{4} it is considered a space-time described by a classical time and a non-commutative space $R^d$ defined by Snyder algebra and the area of the disc and sphere is shown to be
quantized.  Their proof is based on manipulating Snyder algebra so that the eigenvalue problem relates to that of the angular momentum in $ d=2 $ and $ d=3 $.
In this work we use an explicit representation of Snyder algebra, which can be easily extended to the generic case of the sphere $S^d$, and we are able to understand
the structure of the eigenvalues and of the eigenfunctions in an exhaustive way.

This work then relates to our previous results on the quantization of the harmonic oscillator in $ d $ dimensions in the presence of Snyder algebra,
from which we deduce that the quantization of the area of the sphere $S^d$ is obtainable from the spectrum of the harmonic oscillator in a particular limit $ \mu \rightarrow 0 $
( the $\mu$ parameter is introduced in \cite{5} - \cite{6} ).

\section{Area of the sphere in d = 2}

The area of the sphere is in general a function of the radial coordinate. In non-commutative geometry the radial coordinate is replaced by a linear operator acting on an auxiliary Hilbert space. The possible measures of the area of the sphere are identifiable with the eigenvalues of this linear operator. In \cite{4} the eigenvalues of the following operator have been computed in $d=2$ and $d=3$

\beq \widehat{R}^2 \ = \ \sum_{i=1}^d \ ( X_i )^2 \label{21} \eeq

in the presence of Snyder algebra, an important example of non-commutative geometry. In $d=3$ this definition coincides precisely with the area of the sphere.

In this article we generalize the computation of the eigenvalues and eigenvectors of the $ \widehat{R}^2 $ operator to the case of a generic sphere $S_d$ using a particular representation of the Snyder algebra in the compact variable $\rho$:

\beq X_i \ = \ i \hbar \ \sqrt{ \ 1 \ - \ \beta \rho^2 \ } \ \frac{\partial}{\partial \rho_i} \ \ \ \ \ \ \ \ \ p_i \ = \ \frac{\rho_i}{\sqrt{ \ 1 \ - \ \beta \ \rho^2} \ } \ \ \ \ \ \ \ \ \  0 \ < \ \rho^2 \ < \frac{1}{\beta} \label{22} \eeq

The results that we make explicit in this article can be obtained from the case of the harmonic oscillator in $d$ dimensions in the limit of the parameter $ \mu \ \rightarrow \ 0 $ (see \cite{5}-\cite{6}).

We first present the simplest case of the $S_2$ sphere, and then generalize to the case of the generic $S_d$ sphere.

Let's first discuss the spectrum of the operator $ \widehat{R}^2 $ in $d=2$:

\begin{eqnarray} \widehat{R}^2 \ \psi & = & -  \hbar^2 \left[ ( 1 - \beta ( \rho^2_x + \rho^2_y )) \left( \ \frac{\partial^2}{\partial \rho_x^2}  +  \frac{\partial^2}{\partial \rho_y^2}  -  \beta \left( \rho_x \frac{\partial}{\partial \rho_x}  +  \rho_y \frac{\partial}{\partial \rho_y} \right) \ \right) \ \right] \ \psi \nonumber \\
& = & - \hbar^2 \left[ \ ( \ 1 \ - \ \beta \rho^2 \ ) \left( \ \frac{\partial^2}{\partial \rho^2} \ + \ \frac{1}{\rho} \ \frac{\partial}{\partial \rho} \ - \ \frac{l^2}{\rho^2} \ \right)
\ - \ \beta \ \rho \ \frac{\partial}{\partial \rho} \ \right] \ \psi \ = \ R^2 \ \psi \label{23}
\end{eqnarray}

where $l$ is the eigenvalue of the angular momentum in $d=2$. Let $ z = \sqrt{\beta} \ \rho $, the eigenvalue equation becomes:

\beq \left[ ( 1-z^2) \left( \frac{\partial^2}{\partial z^2} \ + \ \frac{1}{z} \ \frac{\partial}{\partial z} \ - \ \frac{l^2}{z^2} \ \right) \ - \ z \frac{\partial}{\partial z} \ \right]
\ \psi \ = \ - \frac{R^2}{\hbar^2 \ \beta} \ \psi \label{24} \eeq

We need to isolate the contribution of the angular part

\beq \psi (z) \ = \ z^l \ P(z) \label{25} \eeq

where $P(z)$ is a polynomial in the variable $z$

\beq \left[ \ ( 1-z^2 ) \frac{\partial^2}{\partial z^2}  \ + \ \left( \frac{1+2l}{z} \ - \ 2 ( 1+l ) z \right) \ \frac{\partial}{\partial z} \ + \ \left( \ \frac{R^2}{\hbar^2 \ \beta} \ - l \
\right) \right] \ P(z) \ = \ 0 \label{26} \eeq

Finally, we pass from the variable $z$ to the variable $x=z^2$ obtaining

\beq \left[ \ x ( 1-x ) \ \frac{\partial^2}{\partial x^2} \ + \ \left( \ ( l+1) \ - \ \left( l + \frac{3}{2} \right) \ x \right) \ \frac{\partial}{\partial x} \ + \ \frac{1}{4} \ \left(
\ \frac{R^2}{\hbar^2 \ \beta} \ - \ l \right) \ \right] \ P(x) \ = \ 0 \label{27} \eeq

a hypergeometric equation with coefficients ($a,b,c$):

\begin{eqnarray}
a & = &  \frac{1}{2} \ \left( \ l + \frac{1}{2} \ + \ \sqrt{l^2+ \frac{R^2}{\hbar^2 \ \beta}+ \frac{1}{4}} \
\right) \nonumber \\
b & = &  \frac{1}{2} \ \left( \ l + \frac{1}{2} \ - \ \sqrt{l^2+ \frac{R^2}{\hbar^2 \ \beta}+ \frac{1}{4}} \
\right) \nonumber \\
c &  = & \ l+1
\label{28} \end{eqnarray}

The general solution is therefore

\beq \psi(z) \ = \  z^l \ {}_2 F_{1} \ \left[ \ \frac{1}{2} \ \left( \ l + \frac{1}{2} \ + \ \sqrt{l^2+ \frac{R^2}{\hbar^2 \ \beta}+ \frac{1}{4}} \
\right) \ , \ \frac{1}{2} \ \left( \ l + \frac{1}{2} \ - \ \sqrt{l^2+ \frac{R^2}{\hbar^2 \ \beta}+ \frac{1}{4}} \
\right) \ , \ l+1 \ ; z^2 \ \right] \label{29} \eeq

Polynomial solutions are obtained when the coefficient $b$ is a negative integer:

\beq b \ = \  \frac{1}{2} \ \left( \ l + \frac{1}{2} \ - \ \sqrt{l^2+ \frac{R^2}{\hbar^2 \ \beta}+ \frac{1}{4}} \
\right) \ = \ - \ n \ \label{210} \eeq

from which we obtain as possible values of the area of the sphere in $ d = 2 $:

\beq R^2 \ = \ \hbar^2 \ \beta \ \left[ \ 4 n^2 + 4 n \left( l + \frac{1}{2} \right) + l \ \right] \label{211} \eeq

To compare with the results of the article \cite{4} we must introduce a new quantum number $ N = 2n + l $ obtaining full agreement

\beq R^2 \ = \ \hbar^2 \ \beta \ [ \ N ( N+1 ) - l^2 \ ] \label{212} \eeq.

Now let's analyze the eigenfunctions in detail. When the coefficient $b=-n$, the coefficient $a=n+l+\frac{1}{2} $ hence the eigenfunction corresponding to the quantum numbers ($n,l$) is

\beq \psi_{n,l} (z) \ = \ z^l \ {}_2 F_1 \ \left( n + l + \frac{1}{2}, - n , l+ 1 ; z^2 \right) \label{213} \eeq

This particular hypergeometric function is nothing more than a Jacobi polynomial with $\alpha=l$ and $\beta=-\frac{1}{2}$:

\beq {}_2 F_1 \ \left( n + l + \frac{1}{2}, - n , l+ 1 ; z^2 \right) \ = \ \frac{ n! \ \Gamma (l+1)}{\Gamma (n+l+1)} \ P_n^{(l,-\frac{1}{2})} \ ( 1-2z^2 ) \label{214} \eeq

We can calculate the normalization of the eigenfunction $\psi_{n,l}(z)$ starting from the known normalization of the Jacobi polynomials:

\beq  \int_{-1}^1  \ ( 1-w )^\alpha \ ( 1+w )^\beta \ P_n^{\alpha,\beta} (w) \ P_m^{\alpha,\beta} (w) \ dw
\ = \ \frac{2^{ \alpha + \beta + 1}}{ 2n + \alpha + \beta + 1 } \ \frac{\Gamma (n+\alpha+1) \ \Gamma (n+\beta+1)}{\Gamma(n+\alpha+\beta+1) \ n! } \ \delta_{n,m} \label{215}
\eeq

With simple steps we get

\beq \int^1_0 \ dz \ z \ (1-z^2)^{-\frac{1}{2}} \ \psi_{n,l} (z) \ \psi_{m,l} (z) \ = \ \frac{1}{4n+2l+1} \ \frac{ n! \ \Gamma (n+ \frac{1}{2}) \ ( \ \Gamma( l+1) \ )^2 }{\Gamma(n+l+1) \  \Gamma(n+l+\frac{1}{2})} \ \delta_{n,m} \label{216} \eeq

Note the factor $(1-z^2)^{-\frac{1}{2}} $ in the normalization due to the choice of the representation (\ref{22}) of the Snyder algebra in the compact variable $\rho$.

\section{General case}

For $d\neq 2$ the linear operator to consider is the following:

\beq \widehat{R}^2 \ = \ - \ \hbar^2 \ \left[ \ ( 1- \beta \rho^2 ) \nabla^2_\rho \ - \ \beta \rho \frac{\partial}{\partial \rho} \ \right] \ = \ R^2 \ \psi \label{31} \eeq

where the operator $ \nabla^2_\rho $ is generally defined by

\beq \nabla^2_\rho \ = \ \frac{1}{\rho^{d-1}} \ \frac{\partial}{\partial \rho} \ \rho^{d-1} \ \frac{\partial}{\partial \rho} \ - \ \frac{l(l+d-2)}{\rho^2} \label{32} \eeq

Let's set $ z = \sqrt{\beta} \rho $ again, the eigenvalue equation to be solved is the following:

\beq \left[ \ ( 1-z^2 ) \ \left( \ \frac{\partial^2}{\partial z^2} \ + \ \frac{d-1}{z} \ \frac{\partial}{\partial z} \ - \ \frac{l ( l+d-2 )}{z^2} \ \right) \ - \ z \frac{\partial}{\partial z} \ + \ \frac{R^2}{\hbar^2 \ \beta} \ \right] \ \psi \ = \ 0 \label{33} \eeq

We isolate the angular factor as before:

\beq \psi(z) \ = \ z^l \ P(z) \label{34} \eeq

where $ P (z) $ is a polynomial in the variable $ z $, from which we obtain:

\beq \left[ \ (1-z^2) \ \frac{\partial^2}{\partial z^2} \ + \ \left( \ \frac{d-1+2l}{z} \ - \ ( d+2l) \  z \ \right) \ \frac{\partial}{\partial z} \ + \ \left( \
\frac{R^2}{\hbar^2 \  \beta} \ - \ l \ \right) \ \right] \ P(z) \ = \ 0 \label{35} \eeq

We set $ x = z^2 $ and we get a hypergeometric equation again

\beq \left[ \ x(1-x) \ \frac{\partial^2}{\partial x^2} \ + \ \left( \ l + \frac{d}{2} \ - \ \left( \ l + \frac{d+1}{2} \ \right) \ x \right) \ \frac{\partial}{\partial x} \ + \
\frac{1}{4} \ \left( \ \frac{R^2}{\hbar^2 \ \beta} \ - \ l \ \right) \ \right] \ P(x) \ = \ 0 \label{36} \eeq

with coefficients

\begin{eqnarray}
a & = &  \frac{1}{2} \ \left( \ l + \frac{d-1}{2} \ + \ \sqrt{l^2 + l(d-2) + \frac{(d-1)^2}{4} + \frac{R^2}{\hbar^2 \ \beta}} \
\right) \nonumber \\
b & = &  \frac{1}{2} \ \left( \ l + \frac{d-1}{2} \ - \ \sqrt{l^2 + l(d-2) + \frac{(d-1)^2}{4} + \frac{R^2}{\hbar^2 \ \beta}} \
\right) \nonumber \\
c &  = & \ l \ + \ \frac{d}{2}
\label{37} \end{eqnarray}

The spectrum follows from the condition that the coefficient $ b $ is a negative integer:

\beq R^2 \ = \ \hbar^2 \ \beta \ \left[ \ 4 n^2 + 4n \ \left( \ l + \frac{d-1}{2} \ \right) + l \ \right] \label{38} \eeq

We introduce again the total quantum number $ N = 2n + l $ and obtain

\beq R^2 \ = \ \hbar^2 \ \beta \ [ \ N( N+ d-1) \ - \ l ( l+ d-2) \ ] \label{39} \eeq

which generalizes in a simple way the result contained in \cite{4} for $ d = 2 $.

For $ l = N $ we obtain the Bekenstein quantization condition of the area of the event horizon:

\beq R^2 \ = \ \hbar^2 \ \beta \ N \ \label{310} \eeq

We compute the eigenfunctions associated with the spectrum (\ref{39})

\begin{eqnarray} \psi_{n,l} (z) & = & z^l \ {}_2 F_1 \left( \ n+l+\frac{d-1}{2}, -n , l+\frac{d}{2} ; z^2 \ \right) \nonumber \\
& = & z^l \ \frac{n! \ \Gamma( l+\frac{d}{2})}{\Gamma( n+l+\frac{d}{2})} \ P_n^{(l+\frac{d}{2}-1,-\frac{1}{2})} \ ( 1-2z^2 )\label{311} \end{eqnarray}

Also in this case the normalization of the eigenfunction $ \psi_{n,l} (z) $ can be calculated with simple steps from that of the Jacobi polynomials:

\beq \int^1_0 \ dz \ z^{d-1} \ ( 1-z^2 )^{-\frac{1}{2}} \ \psi_{n,l} (z) \ \psi_{m,l} (z) \ = \ \frac{1}{4n+2l+d-1} \ \frac{n! \ \Gamma( n+\frac{1}{2}) \ (\Gamma(l+\frac{d}{2}))^2}{\Gamma(n+l+\frac{d}{2}) \ \Gamma(n+l+\frac{d-1}{2})} \ \delta_{n,m} \label{312} \eeq

\section{Conclusions}

In this article the calculation of the area of the sphere $ S_d $ in the presence of the Snyder algebra has been solved exactly and it coincides in every dimension with the Bekenstein conjecture for the area of the horizon of a black hole. This eigenvalue problem can be linked to the case of the harmonic oscillator in the presence of Snyder algebra by means of an appropriate limit.

Note that the eigenfunctions can be connected to tabulated functions, the Jacobi polynomials. It is a remarkable fact that the normalization of eigenfunctions faithfully derives from that of Jacobi polynomials. In particular, to satisfy the orthogonality condition, a non-trivial measure of the integral is required, which can be deduced from the known normalization of Jacobi polynomials with a simple coordinate transformation.

These results encourage us to think that understanding the Snyder space is a necessary step towards the quantization of gravity in ($3+1$) dimensions.

\section{Data Availability}

The data supporting the findings of this study are available within the article [ and its supplementary material].

\end{document}